# A Uniform Algorithm for All-Speed Shock-Capturing Schemes

Xue-song Li

Key Laboratory for Thermal Science and Power Engineering of Ministry of Education, Department of Thermal Engineering, Tsinghua University, Beijing 100084, PR China

Correspondent author: xs-li@mail.tsinghua.edu.cn

**Abstract**
There are many ideas for developing shock-capturing schemes and their extension for all-speed flow. The representatives of them are Roe, HLL and AUSM families. In this paper, a uniform algorithm is proposed, which expresses three families in the same framework. The algorithm has explicit physical meaning, provides a new angel of understanding and comparing the mechanism of schemes, and may play a great role in the further research. As an example of applying the uniform algorithm, the low-Mach number behaviour of the schemes is analyzed. Then, a very clear and simple explanation is given based on the wall boundary, and a concise rule is proposed to judge whether a scheme has satisfied low-Mach number behaviour.

## 1. Introduction

The subject of Computational Fluid Dynamics (CFD) is usually subdivided into research fields for compressible and incompressible flows, which have their own difficulties and solving methods. For the incompressible computation, the one of the most important problem is pressure checkerboard, which motivates development of the momentum interpolation method (MIM) [1-6]. For the compressible computation, the main difficulty is to capture shock, and thus the compressible methods are known as the shock-capturing scheme and the time-marching algorithm for space and time discretization, respectively.

In the past three decades, the shock-capturing scheme has great development, in which three families, i.e. the Roe [7, 8], HLL [9-11], and AUSM [12-14] families, are the most important and widely used. However, these shock-capturing schemes fail to the low-Mach number, i.e. incompressible, flows because of non-physical scale of pressure fluctuation [15]. In order to cure it, many extended schemes are proposed for all-speed flows based on preconditioning technology [15-19] and other new concepts [20-27]. Thus, there are so many schemes, which are proposed based on different ideas, all achieve the same main aims of capturing shock with or without appropriate low-Mach-number behaviour, although they may have different performances of detail properties such as the carbuncle phenomenon and the shock instability [28, 29]. This fact indicates the importance of a uniform algorithm, which can provide a general framework to understand, analyze and improve these schemes. In fact, there are some successful examples although their frameworks are not general enough. For examples, the shock instability of the Roe scheme is cured by comparing it with the HLL-type scheme [29], and the mechanism of the all-speed schemes is discovered by comparing some Roe-type schemes in a uniform framework [27] which is come from Ref. [16].

Followed by the framework of Ref. [27], a more general algorithm is proposed covering all Roe-, HLL-, and AUSM-type schemes at least. It is very simple, has explicit physical meaning, provides another angel of understanding the mechanism of schemes, and may play a great role in the further research of the schemes.

The outline of the present paper is as follows. Section 2 concisely reviews the original form of three families' schemes. Section 3 proposes the uniform algorithm. Section 4 shows the application of uniform algorithm for extending schemes from shock-capturing to all-speed. Finally, Section 5 states the conclusions.

## 2. The Shock-Capturing Schemes

## 2.1 Governing equations

The governing Navier-Stokes (N-S) equations can be written in the Cartesian form:

$$\frac{\partial Q}{\partial t} + \frac{\partial F}{\partial x} + \frac{\partial G}{\partial y} + \frac{\partial H}{\partial z} = \frac{\partial F^v}{\partial x} + \frac{\partial G^v}{\partial y} + \frac{\partial H^v}{\partial z} + S, \qquad (1)$$

where $Q = \begin{bmatrix} \rho \\ \rho u \\ \rho v \\ \rho w \\ \rho E \end{bmatrix}$ is the vector of conservation variables; $F = \begin{bmatrix} \rho u \\ \rho u^2 + p \\ \rho uv \\ \rho uw \\ u(\rho E + p) \end{bmatrix}$, $G = \begin{bmatrix} \rho v \\ \rho uv \\ \rho v^2 + p \\ \rho vw \\ v(\rho E + p) \end{bmatrix}$,

$H = \begin{bmatrix} \rho w \\ \rho uw \\ \rho vw \\ \rho w^2 + p \\ w(\rho E + p) \end{bmatrix}$ are the vectors of Euler fluxes; $F^v$, $G^v$, and $H^v$ are the vectors of viscous fluxes; $S$ is the source term; $\rho$ is the fluid density; $p$ is the pressure; $E$ is the total energy; and $u, v, w$ are the velocity components in the Cartesian coordinates $(x, y, z)$, respectively.

## 2.2 Roe scheme

The classical Roe scheme can be expressed as follows:

$$\tilde{F}_{\frac{1}{2}}^{Roe} = \frac{1}{2}(\bar{F}_L + \bar{F}_R) - \frac{1}{2}R_{\frac{1}{2}}^{Roe}\Lambda_{\frac{1}{2}}^{Roe}\left(R_{\frac{1}{2}}^{Roe}\right)^{-1}\Delta Q, \qquad (2)$$

where

$$\bar{F} = U\begin{bmatrix} \rho \\ \rho u \\ \rho v \\ \rho w \\ \rho H \end{bmatrix} + \begin{bmatrix} 0 \\ n_x p \\ n_y p \\ n_z p \\ 0 \end{bmatrix}, \qquad (3)$$

$$R^{Roe} = \begin{bmatrix} n_x & n_y & n_z & 1 & 1 \\ n_x u & n_y u - n_z & n_z u + n_y & u - n_x c & u + n_x c \\ n_x v + n_z & n_y v & n_z v - n_x & v - n_y c & v + n_y c \\ n_x w - n_y & n_y w + n_x & n_z w & w - n_z c & w + n_z c \\ n_z v - n_y w + \frac{V_M^2}{2}n_x & n_x w - n_z u + \frac{V_M^2}{2}n_y & n_y u - n_x v + \frac{V_M^2}{2}n_z & H - cU & H + cU \end{bmatrix} \qquad (4)$$

$$\Delta Q = Q_R - Q_L, \qquad (5)$$

$$\Lambda^{Roe} = \begin{bmatrix} \lambda_1 & & & & \\ & \lambda_2 & & & \\ & & \lambda_3 & & \\ & & & \lambda_4 & \\ & & & & \lambda_5 \end{bmatrix}, \quad \lambda_1 = \lambda_2 = \lambda_3 = |U|, \quad \lambda_4 = |U - c|, \quad \lambda_5 = |U + c|, \qquad (6)$$

where $c$ is the sound speed, $V_M^2 = u^2 + v^2 + w^2$, $U = n_x u + n_y v + n_z w$ is the velocity normal to the cell interface, and $n_x$, $n_y$, and $n_z$ are the components of the face-normal vector, respectively.

## 2.3 HLL scheme

The HLL-type scheme can be written as follows:

$$\tilde{F}_{\frac{1}{2}}^{\text{HLL}} = \frac{S_R \bar{F}_L - S_L \bar{F}_R}{S_R - S_L} + \frac{S_R S_L}{S_R - S_L} \Delta Q - \delta \frac{S_R S_L}{S_R - S_L} B \Delta Q \qquad (7)$$

where

$$B \Delta Q = \begin{bmatrix} \Delta \rho \\ \Delta \rho u \\ \Delta \rho v \\ \Delta \rho w \\ \Delta \rho E \end{bmatrix} - \frac{\Delta p}{c^2} \begin{bmatrix} 1 \\ u \\ v \\ w \\ H \end{bmatrix} - \rho \Delta U \begin{bmatrix} 0 \\ n_x \\ n_y \\ n_z \\ U \end{bmatrix}, \qquad (8)$$

$$S_R = \max(b^+, 0), \quad S_L = \min(b^-, 0). \qquad (9)$$

For the different version of HLL-type scheme, the definition of $b^+$, $b^-$, and $\delta$ are different. For example, for the HLLEM scheme [10],

$$b^+ = \max\left(U_{\frac{1}{2}} + c_{\frac{1}{2}}, U_R + c_R\right), \quad b^- = \min\left(U_{\frac{1}{2}} - c_{\frac{1}{2}}, U_L - c_L\right), \quad \delta = \frac{c_{0.5}}{0.5\left|b^+ + b^-\right| + c_{0.5}}, \qquad (10)$$

and for the simplest version, which is known as the Rusanov scheme [11],

$$b^+ = |U|_{\frac{1}{2}} + c_{\frac{1}{2}}, \quad b^- = -|U|_{\frac{1}{2}} - c_{\frac{1}{2}}, \quad \delta = 0. \qquad (11)$$

## 2.4 AUSM scheme

As the representation of the AUSM-type scheme, the AUSM+ scheme [13] can be written as follows:

$$\tilde{F}_{\frac{1}{2}}^{\text{AUSM+}} = \frac{\dot{m}_{\frac{1}{2}} + |\dot{m}_{\frac{1}{2}}|}{2} \begin{bmatrix} 1 \\ u \\ v \\ w \\ H \end{bmatrix}_L + \frac{\dot{m}_{\frac{1}{2}} - |\dot{m}_{\frac{1}{2}}|}{2} \begin{bmatrix} 1 \\ u \\ v \\ w \\ H \end{bmatrix}_R + \dot{p} \begin{bmatrix} 0 \\ n_x \\ n_y \\ n_z \\ 0 \end{bmatrix}, \qquad (12)$$

where

$$\dot{m}_{\frac{1}{2}} = \Pi_{\frac{1}{2}} c_{\frac{1}{2}} \begin{cases} \rho_L & \Pi_{\frac{1}{2}} > 0 \\ \rho_R & \text{otherwise} \end{cases}, \qquad (13)$$

$$\dot{p} = f_{p,L}^+\Big|_\alpha p_L + f_{p,R}^-\Big|_\alpha p_R, \qquad (14)$$

$$\Pi_{\frac{1}{2}} = f_{\Pi,L}^+ + f_{\Pi,R}^-, \qquad (15)$$

$$f_{\Pi}^\pm = \begin{cases} \dfrac{1}{2}\left(\tilde{M} \pm |\tilde{M}|\right) & |\tilde{M}| \geq 1 \\ \pm\dfrac{1}{4}\left(\tilde{M} \pm 1\right)^2 \pm \dfrac{1}{8}\left(\tilde{M}^2 - 1\right)^2 & \text{otherwise} \end{cases}, \qquad (16)$$

$$f_{p}^\pm\Big|_\alpha = \begin{cases} \dfrac{1}{2}\left(1 \pm \text{sign}(\tilde{M})\right) & |\tilde{M}| \geq 1 \\ \dfrac{1}{4}\left(\tilde{M} \pm 1\right)^2 \left(2 \mp \tilde{M}\right) \pm \alpha \tilde{M}\left(\tilde{M}^2 - 1\right)^2 & \text{otherwise} \end{cases}, \qquad (17)$$

$$\alpha = \frac{3}{16}, \quad \tilde{M}_L = \frac{U_L}{c_{1/2}}, \quad \tilde{M}_R = \frac{U_R}{c_{1/2}}. \qquad (18)$$

## 3. A Uniform Algorithm for the Shock-Capturing Schemes
## 3.1 A uniform framework

The shock-capturing schemes in the chapter 2 can be generalized as the sum of a central term $\tilde{F}_c$ and a numerical dissipation term $\tilde{F}_d$ as follows:

$$\tilde{F}_{\frac{1}{2}} = \tilde{F}_{c,\frac{1}{2}} + \tilde{F}_{d,\frac{1}{2}}. \tag{19}$$

The central term $\tilde{F}_c$ can be obtained as follows:

$$\tilde{F}_{c,\frac{1}{2}} = \frac{1}{2}\left(\bar{F}_L + \bar{F}_R\right), \tag{20}$$

where $\bar{F}$ is defined in Eq. (3).

Therefore, the difference of the schemes is reflected in the numerical dissipation term $\tilde{F}_d$. In the following sections a uniform algorithm for $\tilde{F}_d$ will be given.

**3.2 Scalar form of the Roe scheme**
Followed by the Ref. [27], the Roe scheme in the section 2.2 can be rewritten as the following scalar form:

$$\tilde{F}_d = -\frac{1}{2}\left\{\xi\begin{bmatrix}\Delta\rho\\ \Delta(\rho u)\\ \Delta(\rho v)\\ \Delta(\rho w)\\ \Delta(\rho E)\end{bmatrix} + \delta p\begin{bmatrix}0\\ n_x\\ n_y\\ n_z\\ U\end{bmatrix} + \delta U\begin{bmatrix}\rho\\ \rho u\\ \rho v\\ \rho w\\ \rho H\end{bmatrix}\right\}. \tag{21}$$

On the right side of Eq. (21), three terms have explicit physical meaning. The first term is the basic upwind dissipation, the second term is a modification to the interface pressure, and the third term is a modification to the interface fluxes. The first and second terms have critical effect on the non-physical behaviour for low-Mach number flows [27], and the third term plays an important role suppressing the pressure checkerboard mode [27] similar to the MIM [6].

For the Roe-type scheme:
$$\xi = \lambda_1 = |U|, \tag{22}$$

$$\delta p = -\frac{\lambda_4 - \lambda_5}{2}c\beta + \left[\lambda_1 - \frac{\lambda_4 + \lambda_5}{2}\right]\left[U\Delta\rho - \Delta(\rho U)\right], \tag{23}$$

$$\delta U = \frac{1}{\rho}\left(\frac{\lambda_4 + \lambda_5}{2} - \lambda_1\right)\beta + \frac{\lambda_4 - \lambda_5}{2\rho c}\left[U\Delta\rho - \Delta(\rho U)\right], \tag{24}$$

where

$$\beta = \frac{\gamma - 1}{c^2}\left[\frac{V_M^2}{2}\Delta\rho - u\Delta(\rho u) - v\Delta(\rho v) - w\Delta(\rho w) + \Delta(\rho E)\right]. \tag{25}$$

It should be noticed that the scalar form, Eqs. (21) – (25), is completely equal to the vector form in the section 2.2 without any assumption.

**3.3 Scalar form for the Roe and HLL schemes with primary variables**
With the assumption as follows:
$$\Delta(\rho\phi) = \rho\Delta\phi + \phi\Delta\rho, \tag{26}$$
where $\phi$ represents one of the fluid variables, and thus $\beta$ in Eq. (25) becomes

$$\beta = \frac{\Delta p}{c^2}. \tag{27}$$

Therefore, Eq. (21) can also be written as the scalar form with primary variables:

$$\tilde{\boldsymbol{F}}_d = -\frac{1}{2}\left\{\xi\begin{bmatrix}0\\\Delta u\\\Delta v\\\Delta w\\\Delta E\end{bmatrix} + \delta p\begin{bmatrix}0\\n_x\\n_y\\n_z\\U\end{bmatrix} + \delta U\begin{bmatrix}\rho\\\rho u\\\rho v\\\rho w\\\rho H\end{bmatrix} + \delta U_\xi\begin{bmatrix}\rho\\\rho u\\\rho v\\\rho w\\\rho E\end{bmatrix}\right\}, \tag{28}$$

where the terms of $\xi$ and $\delta U_\xi$ are divided from the first term in Eq. (21), and the $\delta U_\xi$ term plays a very similar role to the $\delta U$ terms, because they are only a trivial difference in the energy equation. In fact, the dissipation of the energy equation itself plays trivial effect on the low-Mach number behaviour [27].

The terms of $\delta U$ and $\delta p$ can also be subdivided as follows:
$$\delta U = \delta U_p + \delta U_u, \tag{29}$$
$$\delta p = \delta p_p + \delta p_u, \tag{30}$$
$\delta U_p$, $\delta U_u$, $\delta p_p$, and $\delta p_u$ have the physical meaning of pressure-difference and velocity-difference modifications to the interface fluxes and pressure, respectively.

Therefore, for the Roe scheme:
$$\xi = \rho|U|, \tag{31}$$
$$\delta p_p = \frac{U}{c}\Delta p, \tag{32}$$
$$\delta p_u = (c-|U|)\rho\Delta U, \tag{33}$$
$$\delta U_p = (c-|U|)\frac{\Delta p}{\rho c^2}, \tag{34}$$
$$\delta U_u = \frac{U}{c}\Delta U. \tag{35}$$
Considering
$$\Delta\rho = \frac{\Delta p}{c^2}, \tag{36}$$
$\delta U_\xi$ can also be expressed as follows:
$$\delta U_\xi = |U|\frac{\Delta\rho}{\rho} = |U|\frac{\Delta p}{\rho c^2}. \tag{37}$$

The HLL scheme can also be rewritten as the form of Eq. (28), and thus the corresponding terms are given as follows:
$$\xi = \frac{(S_R+S_L)U - 2(1-\delta)S_R S_L}{S_R - S_L}, \tag{38}$$
$$\delta p_p = \frac{S_R+S_L}{S_R - S_L}\Delta p, \tag{39}$$
$$\delta p_u = -\delta\frac{S_R S_L}{S_R - S_L}\rho\Delta U, \tag{40}$$
$$\delta U_p = -\delta\frac{S_R S_L}{S_R - S_L}\frac{\Delta p}{\rho c^2}, \tag{41}$$
$$\delta U_u = \frac{S_R+S_L}{S_R - S_L}\Delta U. \tag{42}$$
Considering Eq. (36):
$$\delta U_\xi = \frac{(S_R+S_L)U - 2(1-\delta)S_R S_L}{S_R - S_L}\frac{\Delta p}{\rho c^2} \tag{43}$$

## 3.4 Extended Scalar form for the AUSM scheme

For expressing the AUSM scheme in the uniform algorithm, Eq. (28) can also be extended as follows:

$$\tilde{F}_{d,\frac{1}{2}} = -\frac{1}{2}\left\{\xi\begin{bmatrix}0\\\Delta u\\\Delta v\\\Delta w\\\Delta E\end{bmatrix}_{\frac{1}{2}} + \delta p_0\begin{bmatrix}0\\n_x\\n_y\\n_z\\0\end{bmatrix}_{\frac{1}{2}} + \delta p_1\begin{bmatrix}0\\0\\0\\0\\U\end{bmatrix}_{\frac{1}{2}} + \delta U_0\begin{bmatrix}\rho\\\rho u\\\rho v\\\rho w\\\rho H\end{bmatrix}_{\frac{1}{2}} + \delta U_1\begin{bmatrix}\rho\\\rho u\\\rho v\\\rho w\\\rho H\end{bmatrix}_R + \delta U_2\begin{bmatrix}\rho\\\rho u\\\rho v\\\rho w\\\rho H\end{bmatrix}_L + \delta U_\xi\begin{bmatrix}\rho\\\rho u\\\rho v\\\rho w\\\rho E\end{bmatrix}_{\frac{1}{2}}\right\}, \quad (44)$$

where $\delta p$ in Eq. (28) is subdivided into $\delta p_1$ and $\delta p_0$ for modifying the interface pressure of the energy equation and other equations, respectively; and $\delta U$ is subdivided into $\delta U_0$, $\delta U_1$, $\delta U_2$ for modifying the interface, right side and left side fluxes, respectively.

For the AUSM+ scheme in the section 2.4, the corresponding coefficients can be expressed as follows:

$$\xi = 0, \ \delta U_0 = 0, \ \delta p_1 = 0, \ \delta U_\xi = 0, \quad (45)$$

$$\delta U_1 = \left(\Pi c - |\Pi c|\right)_{\frac{1}{2}} - U_R, \quad (46)$$

$$\delta U_2 = \left(\Pi c + |\Pi c|\right)_{\frac{1}{2}} - U_L, \quad (47)$$

$$\delta p_0 = \frac{3}{8}\Delta\left(\tilde{M}^5 p\right) - \frac{5}{4}\Delta\left(\tilde{M}^3 p\right) + \frac{15}{8}\Delta\left(\tilde{M} p\right). \quad (48)$$

Similar to Eqs. (29)-(30), $\delta U_1$, $\delta U_2$ and $\delta p_0$ can also be expressed as the sum of the pressure-difference and Mach-difference terms as follows:

$$\delta p_0 = \delta p_{0,p} + \delta p_{0,u}, \quad (49)$$

$$\delta U_1 = \delta U_{1,p} + \delta U_{1,u}, \quad (50)$$

$$\delta U_2 = \delta U_{2,p} + \delta U_{2,u}, \quad (51)$$

where

$$\delta p_{0,p} = \frac{1}{8}\tilde{M}\left(3\tilde{M}^4 - 10\tilde{M}^2 + 15\right)\Delta p, \quad (52)$$

$$\delta p_{0,u} = \frac{15}{8}p\left(\tilde{M}^4 - 2\tilde{M}^2 + 1\right)\Delta\tilde{M}. \quad (53)$$

$$\delta U_{1,p} = 0, \ \delta U_{2,p} = 0, \quad (54)$$

$$\delta U_{1,u} = \left(\sigma - |\sigma|\right) - U_R, \quad (55)$$

$$\delta U_{2,u} = \left(\sigma + |\sigma|\right) - U_L \quad (56)$$

$$\sigma = \begin{cases}\frac{1}{2}(U_R + U_L) - \frac{1}{2}\Delta|U| & |U| \geq c \\ \frac{1}{2}(U_R + U_L) - \frac{1}{2}\tilde{M}^3\Delta U & \text{otherwise}\end{cases} \quad (57)$$

## 4. The Rules for Extending Shock-Capturing to All-Speed Schemes Based on the Uniform Algorithm
### 4.1 The non-physical behaviour of the shock-capturing schemes for the low-Mach number flows

For a long time, it has been qualitatively realized that the shock-capturing schemes suffer from the non-physical behaviour for the low-Mach number flows. By the asymptotical analysis, Ref. [15] gives a theoretical proof to show what the non-physical behaviour means as follows. In the continuous incompressible flows the pressure varies in space asymptotically with the square of the reference Mach number:

$$p(x,t) = P_0(t) + M_*^2 p_2(x,t), \quad (58)$$

which means that both $p_0$ and $p_1$ in the expansion of $p$ are constant in space for the continuous cases. However, the pressure, which is obtained by the shock-capturing schemes, varies asymptotically with the reference Mach number:

$$p(x,t) = P_0(t) + M_* p_1(x,t), \qquad (59)$$

which means that $p_1$ is not a constant in the discretized field by the shock-capturing schemes, and thus lead to non-physical pressure field.

**4.2 An explanation and rule for all-speed schemes**

About the reason of the non-physical behaviour problem, it is usually due to too large numerical dissipation of the shock-capturing scheme, and Ref. [27] explains what part of the dissipation term leads to the problem, i.e., the coefficient of the velocity-difference term in the momentum equation. In this paper, an easy explanation will be given from a new angle, and a clearer and simpler rule will be proposed for the satisfied low-Mach number property of the all-speed schemes.

The Rusanov scheme Eq. (11), which is the simplest version of the HLL scheme, will be disused as example, because it is very simple but typical enough. Its original form of the numerical dissipation term can be rewritten as follows:

$$\tilde{\boldsymbol{F}}_d^{\text{Rusanov}} = -\frac{1}{2}(c+|U|)\begin{bmatrix}\Delta\rho\\\Delta(\rho u)\\\Delta(\rho v)\\\Delta(\rho w)\\\Delta(\rho E)\end{bmatrix}. \qquad (60)$$

For the wall boundary as shown in the Fig.1, in theory the numerical dissipation should be equal to zero, i.e. $\tilde{\boldsymbol{F}}_d = 0$. However, $\tilde{\boldsymbol{F}}_d^{\text{Rusanov}}$ is obviously unreasonable because the velocity gradient is large and the sound speed is nearly a constant. In fact, the numerical dissipation in Eq. (60) is larger than the reasonable value $1/M$ times because the reasonable numerical dissipation should be proportional to the local velocity as in the incompressible solver. Therefore, the non-physical behaviour problem is not surprising because the numerical dissipation is inversely proportional to the local Mach number.

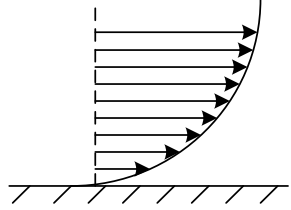

Fig. 1 The wall boundary

For cure the non-physical behaviour problem, Eq. (60) can be modified as:

$$\tilde{\boldsymbol{F}}_d^{\text{P-Rusanov}} = -\frac{1}{2}(c'+|U|)\begin{bmatrix}\Delta\rho\\\Delta(\rho u)\\\Delta(\rho v)\\\Delta(\rho w)\\\Delta(\rho E)\end{bmatrix}, \qquad (61)$$

where $c'$ can be defined as the following simplest form:

$$c' = \min(V, c), \qquad (62)$$

where $V$ is the amplitude of the local velocity:

$$V = \sqrt{u^2 + v^2 + w^2}. \qquad (63)$$

Therefore, Eq. (61) can meet the requirement of the boundary wall, and can satisfy the Eq. (58) which can be deduced easily according the general asymptotical analysis [27].

For Eq. (61), however, another problem occurs that the computation is instability. Then, a global cut-off is adopted for preconditioned HLL-type schemes [18] as follows:

$$c' = \min\left(\max\left(V, V_{\text{ref}}\right), c\right), \tag{64}$$

where $V_{\text{ref}}$ is a global parameter which may has many definitions such as Ref. [18]. Therefore, the numerical dissipation at wall is not zero, and is larger than the reasonable value $M_{\text{ref}}/M$ times. Obviously, for a flow filed in which the Mach numbers in different parts have significant disparity, such as the mixed compressible and incompressible flow, such the global cut-off as in Eq. (64) is obviously unsatisfied. However, all preconditioned shock-capturing schemes [15-19] suffer from this global cut-off problem, although the reason maybe not the same [27].

In order to analyze the behaviour of the Rusanov-type scheme, the expression Eq. (60) can also be rewritten as follows:

$$\tilde{\boldsymbol{F}}_d^{\text{Rusanov}} = -\frac{1}{2}\left\{(c+|U|)\rho\begin{bmatrix}0\\ \Delta u\\ \Delta v\\ \Delta w\\ \Delta E\end{bmatrix} + (c+|U|)\frac{\Delta \rho}{\rho}\begin{bmatrix}\rho\\ \rho u\\ \rho v\\ \rho w\\ \rho E\end{bmatrix}\right\}, \tag{65}$$

Based on the uniform algorithm Eq. (28), Eq. (64) can also be expressed:

$$\delta p_p = 0, \quad \delta p_u = 0, \quad \delta U_p = 0, \quad \delta U_u = 0, \tag{66}$$

$$\xi = \rho(c+|U|), \quad \delta U_\xi = (c'+|U|)\frac{\Delta \rho}{\rho} = (c+|U|)\frac{\Delta p}{\rho c^2}. \tag{67}$$

Then, Eq. (61) without the global cut-off can be rewritten as follows:

$$\xi = \rho\left[\min(V,c)+|U|\right], \quad \delta U_\xi = \left[\min(V,c)+|U|\right]\frac{\Delta p}{\rho c^2}, \tag{68}$$

and Eq. (61) with the global cut-off can also be rewritten accordingly:

$$\xi = \rho\left[\min\left(\max(V,V_{\text{ref}}),c\right)+|U|\right], \quad \delta U_\xi = \left[\min\left(\max(V,V_{\text{ref}}),c\right)+|U|\right]\frac{\Delta p}{\rho c^2}. \tag{69}$$

It can easily noticed that the non-physical behaviour problem is due to $\xi$ in Eq. (67), but not $\delta U_\xi$. $\delta U_\xi$ provides a MIM-type mechanism similar to $\delta U$, more accurate to say, $\delta U_p$, as discussed in the section 3.3.

Eq. (68) cues the non-physical behaviour problem by modifying $\xi$. At the same time, however, it makes $\delta U_\xi$ too small to suppressing the checkerboard, which can be judged by $O(c^{-2})$ order of the coefficient of $\Delta p$ [27], and thus leads to computational instability. Eq. (69) increases $\delta U_\xi$, however, it simultaneously increases $\xi$.

The problem in Eq. (68) and (69) is to modify the $\xi$ and $\delta U_\xi$ simultaneously. A good choice is to treat them respectively as follows:

$$\xi = \rho\left[\min(V,c)+|U|\right], \quad \delta U_\xi = (c+|U|)\frac{\Delta p}{\rho c^2}, \tag{70}$$

where $\delta U_\xi$ is not tends to zero when Mach number tends to zero, which is the inevitably cost for suppressing the checkerboard, but the effect on the accuracy is little fortunately.

In order to validate the above discussion further, the pressure contour is given for the two-dimensional Euler flow past a cylinder with an inflow Mach number of 0.01. Fig. 2 shows that the Rusanov scheme Eq. (60) obtains a solution resembling the Stokes flow, which means full viscous flow rather than Euler flow. As expected, the improved scheme Eq. (70) can converge well and obtain physical solution for the Euler flow.

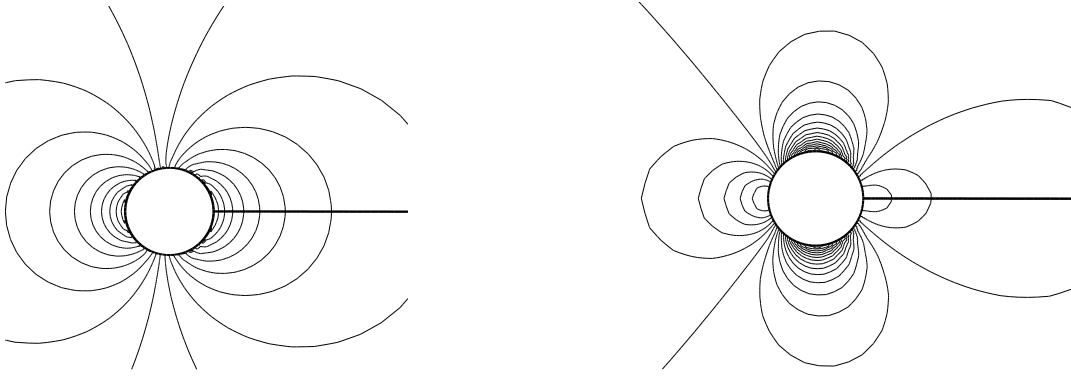

Fig. 2 The solution by Rusanov scheme Eq. (60)    Fig. 3 The solution by improved scheme Eq. (70)

Employing the method given above for the Rusanov scheme, other shock-capturing scheme in the section 3 can also be analyzed similarly. Combining with analysis in Ref. [27], a concise rule can be proposed based on Eq. (28) for satisfied low-Mach number behaviour as follows:

$$\xi = \mathrm{O}(u) \to 0,\ \delta p_u = \mathrm{O}(u\Delta u) \to 0,\ \delta U_p\ \text{or}\ \delta U_\xi = \mathrm{O}\left(c^{-1}\Delta p \sim \Delta p\right) \quad \text{when}\ M \to 0. \tag{71}$$

In detail, $\xi$ and $\delta p_u$ should tends to zero to avoid the non-physical behaviour problem, $\delta U_p$ and $\delta U_\xi$ play the similar role and at least one of them should large enough for suppressing the checkerboard, and $\delta p_p$ and $\delta U_u$ seems trivial for the low-Mach number flows.

The low-Mach number behaviour can be easily uniformly analyzed according to the rule of Eq. (71).

For the non-physical behaviour problem, it is easily due to $\delta p_u$ in Eq. (33) for the Roe scheme, $\xi$ in Eq. (38) and/or $\delta p_u$ in Eq. (40) base on the value of $\delta$ for the HLL scheme, and $\delta p_{0,u}$ in Eq. (53) for the AUSM+ scheme, respectively, because they have the order of $\mathrm{O}(c\Delta u)$.

For the global cut-off problem of the preconditioned technology, it is also due to $\xi$ or $\delta p_u$, which cannot tends to zero although the order $\mathrm{O}(c\Delta u)$ of the original shock capturing schemes is replaced by the order $\mathrm{O}(V_{\mathrm{ref}}\Delta u)$ such as in Eq. (69).

For the checkerboard problem, Roe and HLL schemes have the enough inherent mechanism to suppress it as shown in Eq. (34), and Eq. (41) and/or Eq. (43), respectively. For the AUSM+ scheme, however, such the mechanism is lack, i.e., $\delta U_{1,p} = 0$, $\delta U_{2,p} = 0$ in Eq. (54) and $\delta U_0 = 0$ and $\delta U_\xi = 0$ in Eq. (45). It is the reason why the preconditioned AUSM+ scheme needs an additional terms similar to $\delta U_p$ [19].

## 5. Conclusions

In this paper, a uniform algorithm is proposed, which can express and analyze the shock-capturing schemes and their all-speed extended version from the uniform viewpoint. Therefore, it seems clear and simple. Based on the uniform algorithm, a concise rule is also proposed, which gives criteria to know a scheme whether has the satisfied low-Mach number behaviour and how to improve it.

**Acknowledgments**

This work is supported by Project 51276092 of the National Natural Science Foundation of China.